\documentclass[preprint2]{aastex631}
\begin{document}

\title{Henrietta Leavitt's classical Cepheid relation}

\author[0000-0001-8803-3840]{Daniel Majaess}
\affiliation{Mount Saint Vincent University, Halifax, Nova Scotia, Canada.}
\email{Daniel.Majaess@msvu.ca}

\begin{abstract}
The \citet{lp12} mean period-magnitude relation for SMC Cepheids exhibits correlated residuals relative to modern Gaia $B_G$ and Johnson $V$ photometry ($r_{\delta V/\delta L12}\approx0.76$).  Therefore dispersion inherent to Leavitt's photographic photometry relation is not solely random scatter ($\sigma_{L12,B_G,V}=0.22\pm0.03,0.25\pm0.04,0.20\pm0^{m}.03$), and is potentially consistent with the partial encoding of combined effects from the elongated SMC, classical Cepheid instability strip, differential reddening, etc. OGLE data suggest that $\gtrsim870$ SMC variables identified by \citet[][Table 2]{lea08} are classical Cepheids ($\gtrsim 90$\%), and the preliminary unphased Table 2 photometry likewise features correlated residuals relative to contemporary observations ($r_{\delta B_G/\delta L08-T2}\approx0.54$).  Table 6 data in \citet[][$n=16$]{lea08} define a highly linear Leavitt law over an extensive baseline ($r\approx-0.98$), but constitute a smaller sample than the later 1912 work ($n=25$). 
\end{abstract}

\section{Introduction}

\begin{figure*}[t]
\begin{center}
 \includegraphics[width=1.05\columnwidth]{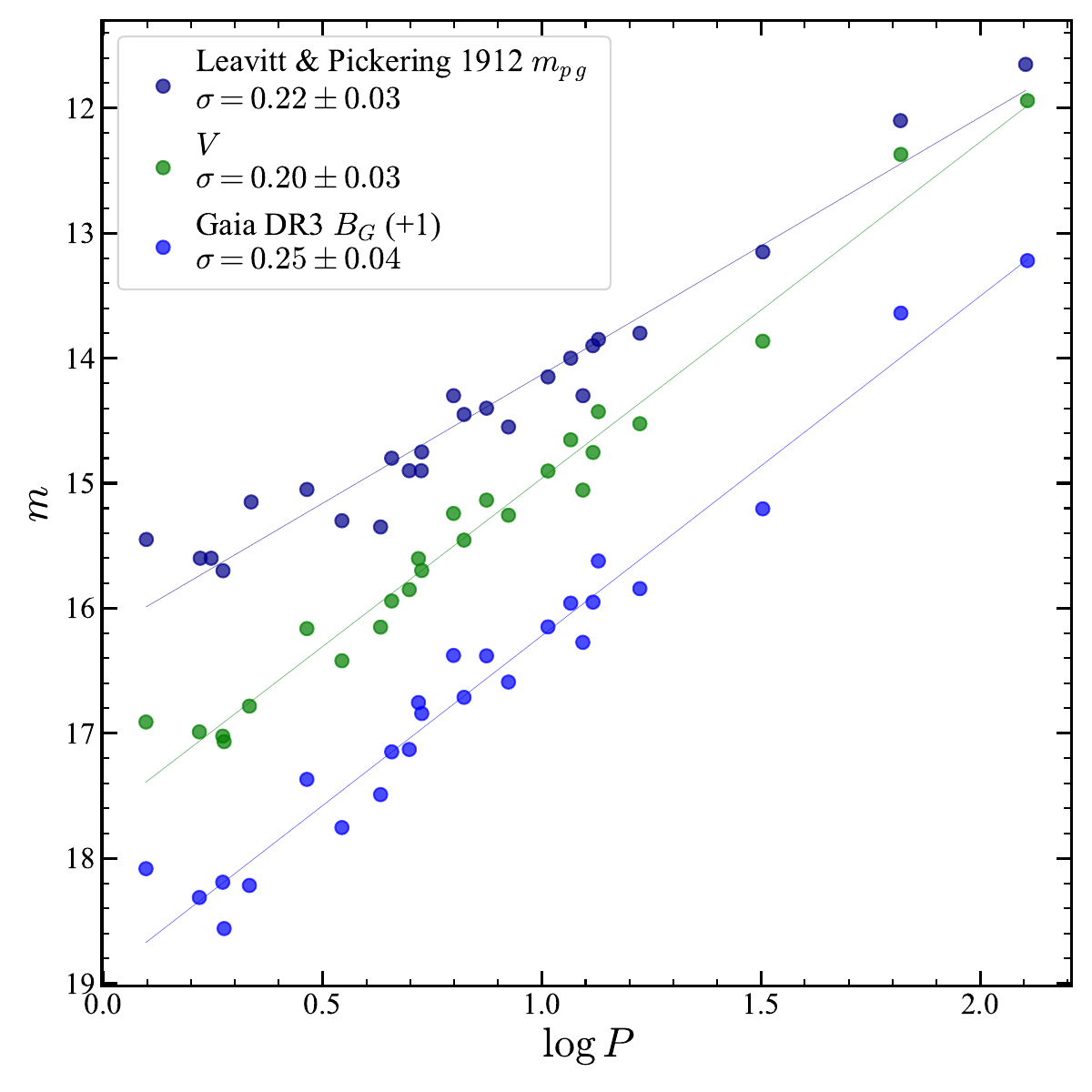}
 \includegraphics[width=1.05\columnwidth]{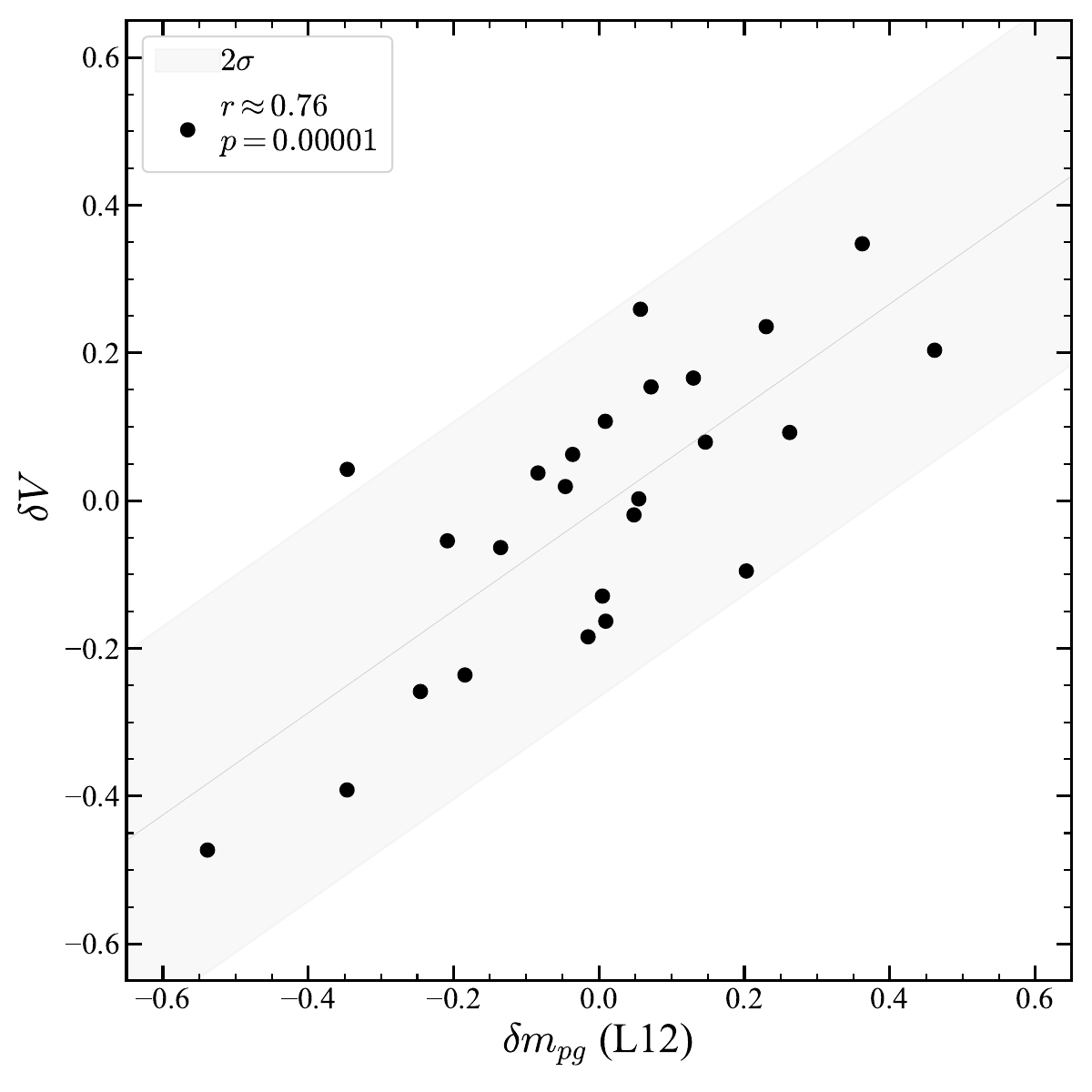}
  \caption{Correlated residuals emerge between the mean \citet{lp12} period-magnitude relation (photographic, $m_{pg}$), Gaia $B_G$, and Johnson $V$ ($n=25$, e.g., $r_{\delta V/ \delta L12}\approx 0.76$, shaded region is $\pm 2\sigma$ about the fit). That suggests various effects may be embedded within Leavitt's photographic photometry.}
 \label{fig-pm}
\end{center}
\end{figure*}

Complexities inherent to the classical Cepheid period-magnitude relation remain actively debated a century after Leavitt's discovery \citep[e.g.,][]{bon24}. Simultaneously, continued insight is emerging regarding Leavitt's historical work \citep[e.g., Fig.~3 in][]{bre25b}.  

In this study, the mean photographic \citet{lp12}\footnote{The opening passage of the work notes, \textit{`The following statement regarding the periods of 25 variable stars in the Small Magellanic Cloud has been prepared by Miss Leavitt.'}} period-magnitude data are shown relative to modern Gaia $B_G$ and Johnson $V$ equivalents (Fig.~\ref{fig-pm}), where a less indirect comparison between mean single optical passbands is preferred when contrasting observed dispersions. Importantly, attention is drawn to the unveiled residuals, which are subsequently investigated within a Pearson correlation framework. Furthermore, period-magnitude data in the earlier \citet[][Table 6]{lea08} work is quantitatively assessed, and the number of classical Cepheids in their Table 2 listing of variables is estimated.

Historically, for example, Leavitt's findings led Hertzsprung to infer the SMC distance, and aided Hubble to discern that the Milky Way was not the Universe.  Leavitt's work was subsequently leveraged in part to help discover an expanding Universe.  As a result Leavitt is honored by the naming of asteroid 5383 Leavitt, a Lunar crater, Leavitt variables, and the classical Cepheid period-magnitude relation is referred to as Leavitt's law, etc. \citep[e.g.,][]{gri85,mad09}.  Leavitt was likewise for example lauded by a US president, and her life is described in dedicated books \citep[e.g.,][]{joh05} and a popular semi-fictional theatre play (\textit{`Silent Sky'} by L. Gunderson).

\section{Analysis}
\label{sec:analysis}
The \citet[][]{lp12} data were cross-matched to Gaia $B_G$ ($BPmag$ from VizieR catalog I/355/gaiadr3), whereby the latter's median number of epochs is 44.  The photographic photometry was likewise compared to the $V$-band \citet{sos15} Cepheid database.  The two brightest stars (HV 821 and HV 824) lacked $V$-photometry in \citet{sos15}, and consequently the magnitudes stem from \citet{mad85} \citep[see also][]{mw79}. Regarding HV 821 \citet[][]{lea08} notes, \textit{`89 observations ... give an average deviation from the light curve of only six hundredths of a magnitude.'}  \citet[][]{tur10b} achieved similar uncertainties when studying century old Harvard photographic plates of separate Galactic stars (e.g., HDE 344787), after years of practice (priv. communication).  

The \citet{lp12} mean magnitudes were deduced here using lightcurve maxima and minima provided in their Table 1, whereas for example the \citet{sos15} equivalent is tied to intensity weighted mean magnitudes.  Key findings remain relatively unaffected if \citet{sos15} magnitudes are similarly determined from maxima and minima.  

\citet{bre25b} concluded that the \citet{lp12} period determinations are highly similar to those established by \citet[][]{sos15}, and pertinently demonstrated that a subset of the lightcurves are likewise comparable (their Fig.~3). Indeed, an inspection of the periods reveals that ${\log{P}}_{OGLE}=(1.0007\pm0.0002){\log{P}}_{L12}-(0.0007\pm0.0002)$ without accounting for Cepheid period changes, and for information on the latter topic see \citet{tur06}.  The longest-period target may not be a canonical Cepheid \citep[HV 821, $P\simeq 128^d$,][]{bt97}, and/or adheres to a slope discontinuity \citep[e.g., $\log{P}\gtrsim2$,][their Fig.~3]{per04}.  On the opposing end the shortest-period Cepheids show signs of crowding and the \citet{san88} bias.  Near the detection limit there is preferential sampling of the brighter hot-edge of the instability strip. There is also a break in the SMC period-magnitude relation near $\log{P}\approx0.4$
\citep[e.g.,][]{uda99,bha16b}.  In sum, there are SMC slope discontinuities (e.g., $\log{P}\approx0.4$ and $\log{P}\approx2$), and see also \citet{pg66}.

\begin{figure}[t]
\begin{center}
 \includegraphics[width=1.0\columnwidth]{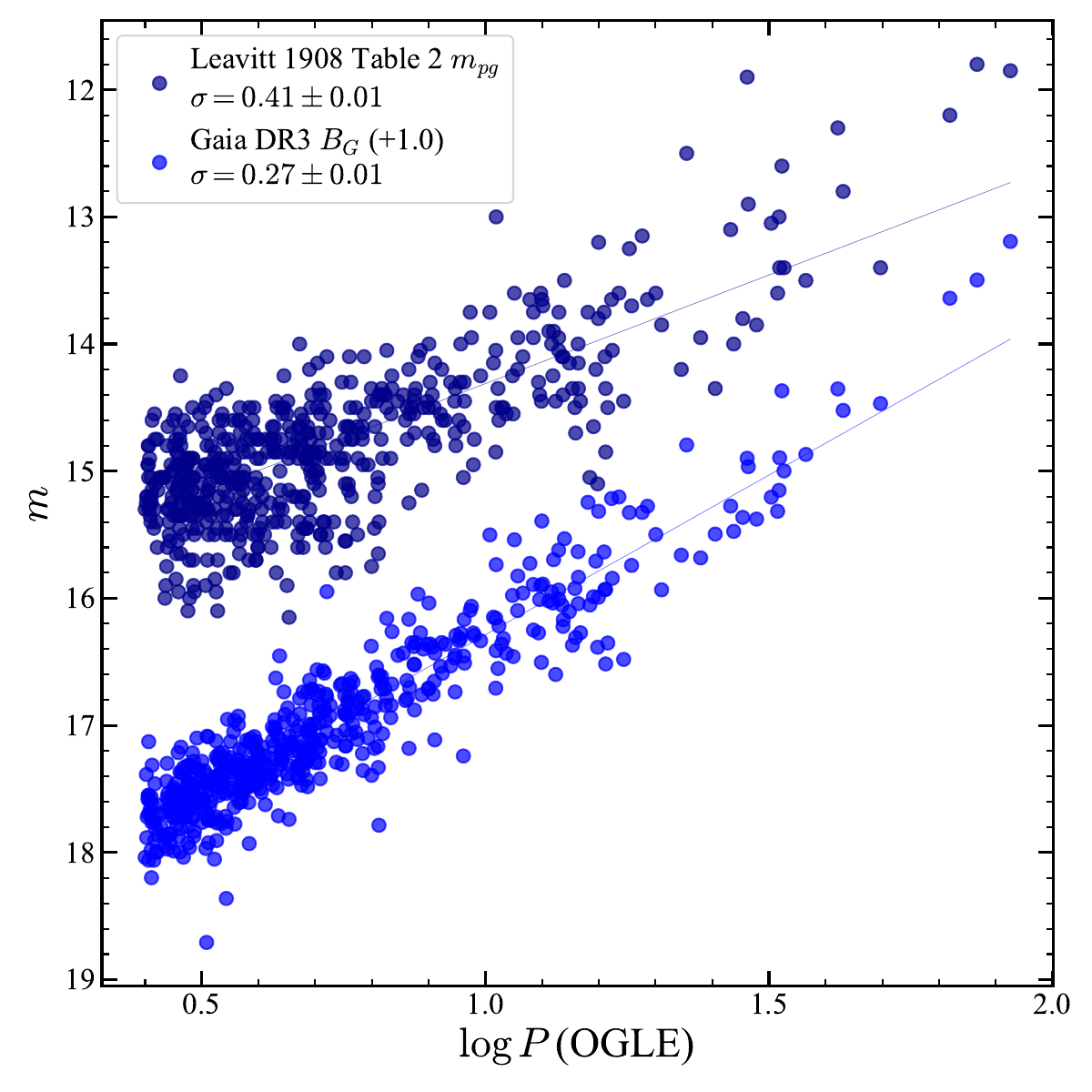} \\
 \includegraphics[width=1.0\columnwidth]{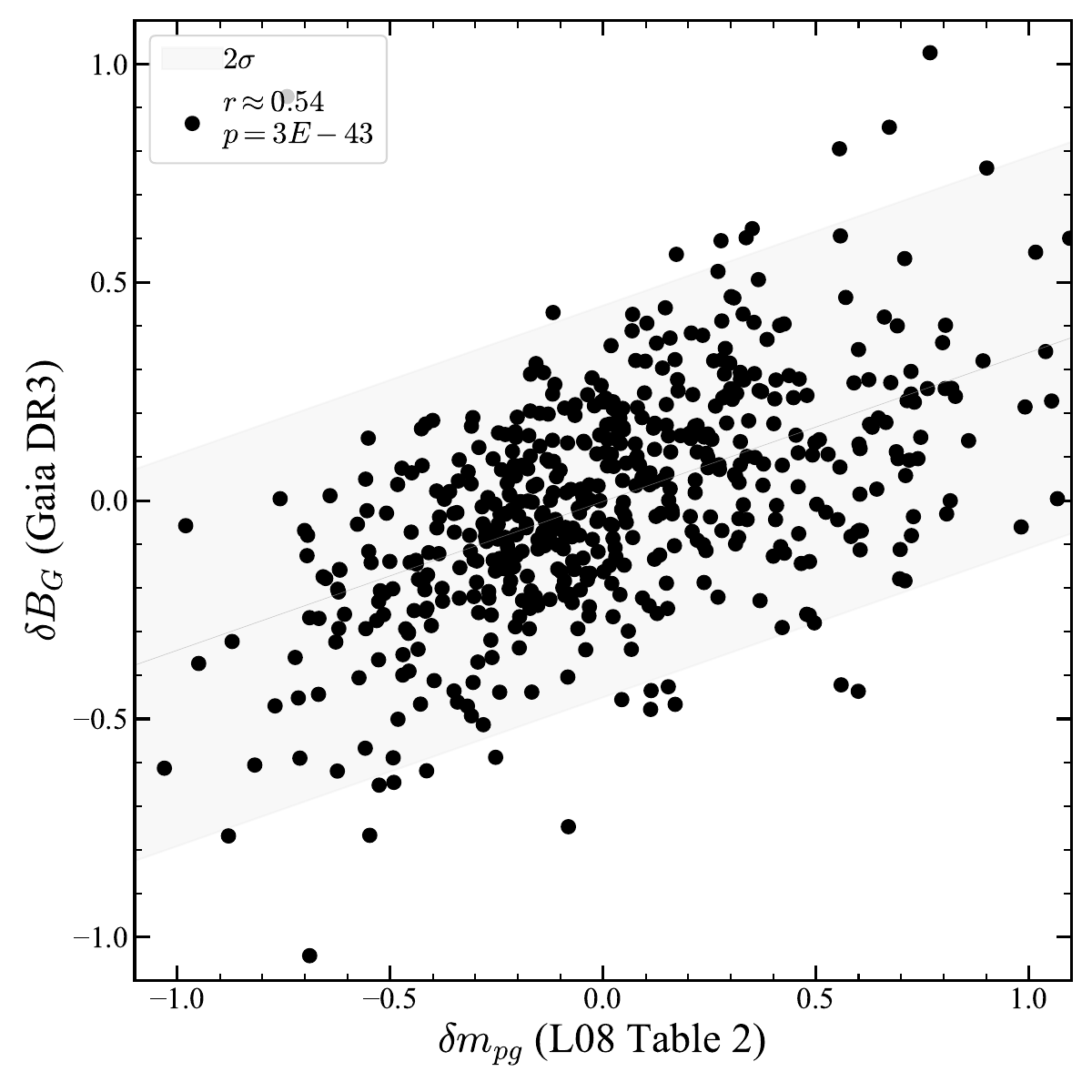} 
  \caption{Correlated residuals exist ($r_{\delta B_G/\delta L08-T2}\approx0.54$) amongst $0.4<\log{P}<2$ fundamental mode classical Cepheids in the unphased \citet{lea08} Table 2 and Gaia $B_G$ datasets (i.e., not determined from lightcurves).  Pulsation periods and classifications are from OGLE.}
 \label{fig-cor2}
\end{center}
\end{figure}

Fig.~\ref{fig-pm} conveys the \citet{lp12} period-magnitude observations, in concert with Gaia $B_G$ and $V$ for overlapping stars.  Robust fitting (Huber) was used to establish linear relations throughout this work.  Gaia $B_G$ data were offset by +1 mag to facilitate visual discernment.  The dispersions are $\sigma_{L12,B_G,V}=0.22\pm0.03,0.25\pm0.04,0.20\pm0^{m}.03$ ($n=25$), yet pertinent considerations complicate a dispersion comparison.  First, the effective wavelength of the photographic plates could be near Johnson B \citep{tan13}, or lie between Johnson B and V \citep[][and amplitude arguments therein partly linked to \citealt{ks09}]{bre25b}.  Consequently, the instability strip should be broader (larger dispersion) for the \citet{lp12} photometry relative to $V$ \citep[Fig.~4 in][]{mf91}.  Expectations are complicated by non-linear and color-term effects endemic to photographic plates, and slope and dispersion changes that are dependent on parameters including brightness (e.g., compression).  Dereddened Johnson-standardized OGLE SMC slopes are $\alpha_{B_0,V_0}=-2.207\pm0.053,-2.572\pm0.042$ mag dex$^{-1}$ \citep[][their Table 1]{uda99}, and to generally match their bounds slopes in Fig.~\ref{fig-pm} were recalculated for $0.4<\log{P}<2$ ($\alpha_{L12,B_G,V}=-2.26\pm0.14,-2.71\pm0.17,-2.76\pm0.15$ mag dex$^{-1}$).   \citet{bre25b} similarly obtained $\alpha_{L12}=-2.19\pm0.11$ mag dex$^{-1}$ ($0.5<\log{P}<1.5$), and consider discussion on photographic plates therein.

Residuals from \citet{lp12} $m_{pg}$ and Johnson $V$ are shown in Fig.~\ref{fig-pm}, and adhere to a Pearson correlation of $r_{\delta V/\delta L12}\approx0.76$ ($p<0.001$, $n=25$).  \citet{lp12} $m_{pg}$ and Gaia $B_G$ residuals exhibit a correlation of $r_{\delta B_G/\delta L12}\approx0.66$.  Restricting the sample to $0.4<\log{P}<2$ increased the correlated residuals ($r_{\delta V/ \delta L12}$ and $r_{\delta B_G/ \delta L12}\approx0.8$).  Hence photometry extracted by Leavitt from photographic plates could be partially tracking combined shifts consistent with the SMC morphology\footnote{For broader context regarding SMC structure see \citet{sco16} and \citet{rip17}, and for historical context \citet[][and references therein]{pg66}.} \citep[see also \S 4.1 in][who approached matters from a separate viewpoint]{bre25b}, classical Cepheid instability strip, differential reddening, etc. The impetus of this study is to highlight that additional information is embedded within Leavitt's photometry, and future efforts can assess relative contributions of the aforementioned effects and the impact of culling potentially contaminated and non-canonical Cepheids at the boundaries.

\begin{figure}[t]
\begin{center}
 \includegraphics[width=1.0\columnwidth]{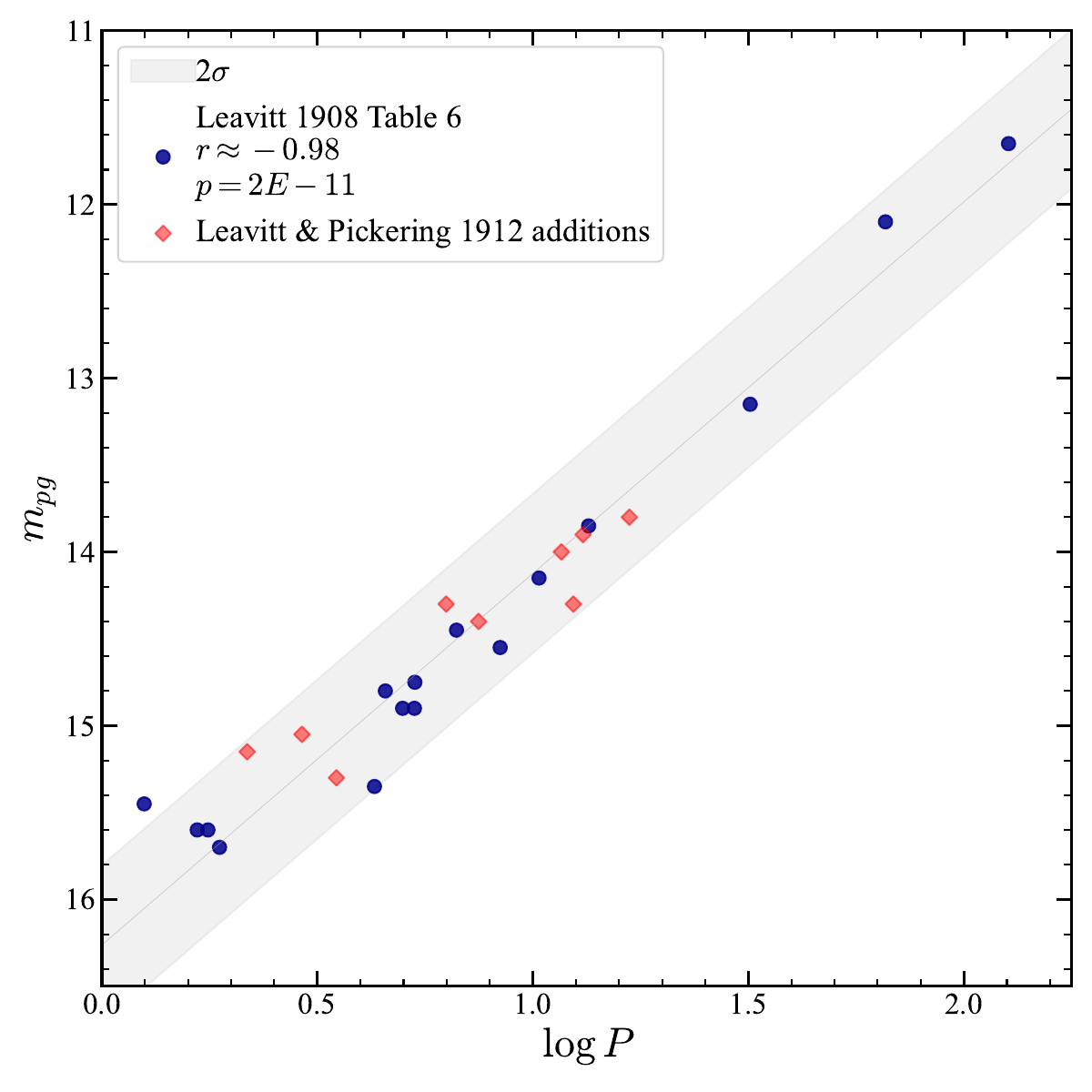}
  \caption{The \citet[][$n=16$]{lea08} Table 6 dataset displays a highly linear relation with expansive coverage ($r\approx-0.98$, shaded region is $\pm 2\sigma$ about the 1908 results), though constituting a smaller sample relative to \citet[][$n=25$]{lp12}.  New datapoints added by \citet{lp12} are represented by red diamonds.}
 \label{fig-1908}
\end{center}
\end{figure}

\citet{lea08} noted that the majority of variables assessed in Table 6 were of the cluster variable type\footnote{Later confirmed to constitute several subclasses.} ($n=16$), whereas the following was remarked regarding variables in their Table 2 ($n=969$), \textit{`While, therefore, the light curves thus far obtained have characteristics to which the majority of the variables will probably be found to conform, no inference can be drawn with regard to the prevalence of any particular type, until many more of the periods have been determined.'} Naturally, some magnitude differences exist between the preliminary \citet{lea08} Table 2 data, and the phased samples that are Table 6 therein and Table 1 in \citet{lp12} (i.e., maximum and minimum derived from the lightcurve). SMC variables from that Table 2 were cross-referenced with the OGLE catalog, and $\gtrsim 90$\% are classical Cepheids ($n\gtrsim870$) \citep[see also][]{pg66}.  The analysis relied on HV stars listed in the OGLE VizieR catalog (OName column), and for entries without HV designations the CDS Sesame Name Resolver was utilized ($r<3$").  Spurious coordinates can occur, and thus the quantity cited above is merely an approximate first-order estimate. 

Next, since the vast majority of Table 2 stars in \citet{lea08} lacked periods, $0.4<\log{P}<2$ fundamental mode OGLE pulsation periods were paired with Table 2 and Gaia $B_G$ magnitudes (brightest source $r<3$" from the CDS HV coordinates), and the ensuing dispersion and correlated residuals are $\sigma_{L08-T2,B_G}=0.41\pm0.01,0.27\pm0^{m}.01$ and $r_{\delta B_G/\delta L08-T2}\approx0.54$, respectively ($n=553$, 4 stars removed with a $5\sigma$ clip, Fig.~\ref{fig-cor2}).  Dispersion for the preliminary \citet{lea08} Table 2 observations is expectedly larger than both the phased Table 6 and \citet{lp12} results, because the latter two possess more robustly determined magnitudes.

Lastly, \citet{lp12} stated that, \textit{`In H.A.~60, No.~4, attention was called to the fact that the brighter variables have the longer periods, but at that time it was felt that the number was too small to warrant the drawing of general conclusions.'}  Those $16$ variables in Table 6 of \citet{lea08} are presented in Fig.~\ref{fig-1908}, and span a sizable baseline.  The relation is highly linear ($r\approx-0.98$, $p<0.001$), thereby laying the groundwork for the subsequent \citet[][$n=25$]{lp12} finding, which relied on larger statistics. Table 6 in \citet[][$n=16$]{lea08} was compared to Table 1 in \citet[][$n=25$]{lp12}, and analogous parameters were confirmed. 

\section{Conclusion}
The \citet{lp12} mean period-magnitude relation for SMC classical Cepheids displays correlated residuals relative to Gaia $B_G$ and $V$ (Fig.~\ref{fig-pm}, and see also Fig.~\ref{fig-cor2} regarding the preliminary \citealt{lea08} Table 2), thereby indicating that Leavitt's photographic observations were not exclusively random scatter about the mean relation. The modern dispersions compared to \citet{lp12} are seemingly similar, yet there are important caveats hindering an unambiguous comparison (\S \ref{sec:analysis}, e.g., $\lambda_{\rm eff, L12}$, photographic plate non-linearity). The vast majority of variables in Table 2 of \citet{lea08} are classical Cepheids.  The earlier phased \citet[][Table 6]{lea08} dataset exhibits a substantial spread in period, and is characterized by a highly linear relation (Fig.~\ref{fig-1908}, $r\approx-0.98$). 

Looking forward, there is evidence that Leavitt may have discovered classical Cepheids in star clusters.  Additional research on that subject is desirable, in tandem with an enhanced evaluation of sources associated with the correlated residuals.\\

\textit{Acknowledgments:} This research relied on initiatives such as CDS, NASA ADS, arXiv, OGLE, Gaia, PHaEDRA. Thanks to kind staff (Alison Doane)\footnote{\url{https://cfa.harvard.edu/news/memoriam-alison-doane}} who welcomed the author when visiting Harvard in 2006, where photographic lightcurves for a subsample of Galactic Cepheids were established.  Access to VizieR and python tools (e.g., OCR) was facilitated by vibe coding in Gemini 3.

\bibliography{article}{}
\bibliographystyle{aasjournal}

\end{document}